# Integrating market sentiments for stock price prediction: A comparative study of Bi-LSTM and multilayer perceptions

Ahmet Akusta [1]

[1] Konya Technical University, Rectorate, Konya, Türkiye

**Abstract** This study investigates the integration of sentiment analysis with machine learning models to forecast stock price movements using the Nvidia Corporation as a case study. Sentiment scores were derived from Nvidia-related financial news headlines using two sentiment analysis tools: FinBERT, a domain-specific tool, and TextBlob, a general-purpose tool. These scores were integrated into predictive frameworks based on bidirectional long short-term memory (Bi-LSTM) networks and multilayer perceptrons (MLPs) developed alongside historical stock price data. This study assesses the predictive performance over the entire observation period and across distinct market phases: bullish, stagnation, and strong bullish conditions. The findings indicate that sentiment features enhance predictive accuracy in specific contexts, particularly during stagnation phases, with TextBlob demonstrating superior performance to FinBERT in specific scenarios. In addition, Bi-LSTM models exhibit consistently superior performance in capturing temporal dependencies compared to MLPs. However, the impact of sentiment features diminished during strongly directional trends, such as those observed in strong bullish markets. The combination of FinBERT and TextBlob in the same dataset allows for a dual-perspective approach to sentiment analysis, thereby providing new insights into the dynamic relationship between market sentiment and stock price behavior. This research contributes to the existing literature on applying sentiment analysis to financial forecasting by advancing the integration of complementary sentiment tools and phase-specific evaluations.

**Keywords** Bi-LSTM · Multilayer perceptrons · Sentiment analysis · Machine learning

Corresponding author: Ahmet Akusta ahmetakusta@hotmail.com

# 1. Introduction

Financial markets are complex systems driven by various psychological and external factors that contribute to fluctuations in stock prices and market trends. Key drivers of market direction include investor sentiment, news, and current events, which influence investor ideology, perspectives, and predictions about current or future trading. Investor sentiment, a psychological factor reflecting investors' mood or attitude toward a particular security or market (Ma et al., 2021), is often a precursor to market changes or price shifts. These sentiments are influenced by daily news and events, causing market movements either upward or downward (Karimova & Rakhmetulayeva, 2023). To enhance the analysis of market data, leveraging technology is essential for understanding the interplay between information and sentiments.

In today's digital era, with the proliferation of social media and online platforms, financial news significantly impacts stock prices as investors often react to current events and breaking news (Saxena et al., 2023). Sentiment analysis, which involves extracting and quantifying opinions or sentiments using Natural Language Processing (NLP), plays a vital role in addressing challenges in financial markets. The proposed model enables systematic analysis of subjective information within datasets, providing insights into public sentiment and its impact on financial trends.

Applying sentiment analysis to financial news from various sources, such as social media posts and news headlines, allows the identification of patterns in capital market movements. This information is highly valuable for decision-making and trading (Varija & Hegde, 2024). By gauging public sentiment toward a particular company, investors and analysts can make informed predictions about market behavior (Bouktif et al., 2020). Sentiment analysis, therefore, may help bridge the gap between financial theory and market psychology, potentially becoming an important tool in modern investing. As this technology evolves, combining it with machine learning models can improve the accuracy of stock market predictions (Kazemian et al., 2016).

Machine learning (ML) has revolutionized financial market analysis, providing tools to process large volumes of data and identify patterns. Among these tools, Recurrent Neural Networks (RNNs) are particularly effective in capturing sequential data patterns, making them ideal for financial time series analysis (Bouktif et al., 2020). RNNs, operating through recurrent loops, process historical and new data simultaneously, making them well-suited for predicting stock prices (Kazemian et al., 2016). Integrating RNNs with sentiment analysis offers a powerful framework for combining quantitative trends with qualitative market insights, which can significantly enhance financial modeling (Varija & Hegde, 2024).

Recurrent Neural Networks (RNNs) are well-suited for sequential data tasks, as they maintain an internal memory that updates at each time step based on the current input and previous hidden states. However, traditional RNNs face challenges, particularly with long-range dependencies, due to the vanishing gradient problem (Zhang et al., 2022). This problem makes it difficult for RNNs to learn from long sequences, leading to a tendency to forget earlier inputs in tasks that require long-term dependencies (Alam et al., 2022). To address these challenges, more advanced RNN variants, such as Long Short-Term Memory (LSTM), Bidirectional LSTM (BiLSTM), and Gated Recurrent Units (GRU), offer enhanced capabilities. For example, BiLSTM processes sequences in both forward and backward directions, capturing dependencies from the past and future, thus improving performance in tasks like network intrusion detection (Zhang et al., 2023). In addition, BiLSTM has demonstrated superior prediction accuracy in various applications, outperforming traditional

RNNs and LSTMs in tasks such as stock price prediction, where it achieves lower error rates and better fitting between predicted and actual values (Yang et al., 2023).

The primary motivation of this study is to address critical gaps in financial forecasting, specifically focusing on the effectiveness of sentiment analysis tools and machine learning architectures in predicting market behavior. First, sentiment analysis has proven valuable in market prediction; however, the effectiveness of specific tools like FinBERT (financial-specific) versus TextBlob (general-purpose), still needs to be explored. Second, a direct comparison of machine learning architectures—such as Bidirectional Long Short-Term Memory (Bi-LSTM) networks and Multi-Layer Perceptrons (MLPs)—has not been sufficiently studied for integrating sentiment and historical data. Lastly, there is a need to evaluate the performance of these methods across different market phases (e.g., bullish, stagnation, and strong bullish), offering insights into the dynamic role of sentiment in financial markets.

FinBERT and TextBlob were chosen as the sentiment analysis tools due to their complementary capabilities in capturing nuanced market sentiment, especially in diverse financial contexts. FinBERT, a model pretrained on financial texts, can extract domain-specific sentiments from news articles and market reports, providing precise insights critical for decision-making in volatile markets (Eik Den Yeoh et al., 2023). The ability to process sentiment in financial texts aligns well with the sensitivity of stock prices to market news, making it particularly effective in phases where investor sentiment drives price movements. In contrast, TextBlob offers a lightweight and versatile approach to general sentiment analysis, serving as a complementary tool to quickly and effectively gauge public sentiment (Gujjar & Kumar, 2020). This dual usage ensures a comprehensive sentiment analysis approach that combines domain-specific depth and general applicability.

Bi-LSTM networks and multilayer perceptrons (MLPs) were selected as machine learning models to address different aspects of stock price prediction. Bi-LSTM's ability to model long-term dependencies in time-series data makes it suitable for analyzing stock price trends influenced by sentiment changes over time (Kurniawan et al., 2023). Its bidirectional processing capability allows it to capture past and future dependencies, which are valuable in scenarios like stagnation or transitional market phases. MLPs, with their simplicity and computational efficiency, provide a reliable baseline for initial predictions and are particularly useful in less complex market conditions. By integrating Bi-LSTM for detailed temporal modeling and MLPs for efficient predictions, the proposed approach ensures robustness and adaptability across varying market conditions, thereby enhancing predictive accuracy in different phases of market behavior.

This study evaluates the integration of market sentiment analysis with stock price predictions for Nvidia Corporation, a major technology firm. By combining historical stock data with sentiment scores derived from news headlines, this study explores the predictive power of Bi-LSTMs and MLPs. The selected models and sentiment analysis tools were applied by processing Nvidia-related news headlines using FinBERT and TextBlob to obtain sentiment scores, which were then incorporated as features into the predictive models. Bi-LSTMs were employed to capture sequential patterns in the time-series data. At the same time, MLPs served as a comparative baseline to assess the impact of model complexity on prediction performance. In addition, this study compares FinBERT and TextBlob sentiment analysis tools to assess their impact on model performance. This comparative framework provides actionable insights for financial analysts and investors, emphasizing the broader utility of sentiment analysis in enhancing financial forecasting models.

## 2. Literature review

This literature review acknowledges the transformative impact of machine learning (ML) and deep learning (DL) techniques in stock price prediction, particularly when augmented with sentiment analysis. This introduction sets the stage for a detailed exploration of how integrating these advanced computational methods with sentiment analysis enriches the prediction of stock market movements.

(Kalyani et al., 2016) examined the predictive capability of financial news articles and presented initial exploratory work in this field. This work used Random Forest and SVM models to achieve predictive accuracy that was well above random and paved the way for increased work in the area. This foundational work was further developed by Bharathabau et al. (2023), who incorporated technical indicators into sentiment analysis using machine learning. They were one of the pieces of work highlighting the potential of merging sentiment analysis with technical indicators and, by doing so, potential refining decisions made by traders.

Wang (2023) followed this work by analyzing multilingual tweets and discovered that global sentiment had tracked stock prices, suggesting the usefulness of broadening the analysis toward linguistic diversity. Similarly, (Štrimaitis et al., 2021) investigated sentiment analysis in the financial sector employing the Lithuanian language. They highlighted the challenges and potential of applying this machinery to various languages practically without NLP tools. In addition, they highlighted the importance of embracing linguistic diversity in sentiment analysis, and they underscored the challenges and potential benefits of applying NLP tools to multiple languages.

The next progression of model sophistication is explored through the work of (Liu et al., 2023), who used FinBERT in conjunction with ensemble SVM techniques and demonstrated the efficacy of combining multiple ML models. This is followed by Kaeley et al. (2023), who demonstrated the superiority of transformer models over RNNs in capturing the dynamic nature of stock market sentiment. Together, these studies highlight the progression of computational techniques in sentiment analysis from basic ML models to more advanced, superior mechanisms in terms of accuracy.

Cui et al. (2022) proposed a novel approach that integrates Text-rank-based text summarization and sentiment analysis to improve stock prediction. Combining text summarization and prior knowledge derived from sentiment analysis requires higher efficiency when handling a large volume of textual data. Another way to optimize sentiment analysis is by optimizing the embedding techniques (Yekrangi & Nikolov, 2023).

Sidek et al. (2023) and Ridhawi and Osman (2023) performed sentiment analysis on news headlines and social media. These studies demonstrate an expanding range of sentiment sources—from localized news to global social media platforms—highlighting how different sentiment inputs can affect stock market predictions.

(Liu et al., 2023) and Xiao et al. (2023) explored the connection between social media sentiment, quant data, and genetic algorithms and the correlation between the temporal characteristics of sentiment. These studies provide insights into various possibilities for combining sentiment analysis with computational methods, such as genetic algorithms, to explore diverse market sentiment characteristics.

Araci (2019) proposed FinBERT, which is used to customize NLP models for financial sentiment analysis. This development could be significant because it introduces a specialized tool that can interpret financial language with greater accuracy, potentially addressing some limitations of general-purpose sentiment analysis tools.

With the severe challenge of the COVID-19 pandemic for the world, Costola et al. (2023) and Sharaf et al. (2023) used the latest updates and news as input to assess the impact on the stock market. These studies suggest that sentiment analysis can be adapted to account for environmental and global changes, such as the COVID-19 pandemic.

Despite these advancements, areas in the literature that could benefit from further exploration are still needed. Previous studies, such as those by Kaeley et al. (2023), have emphasized the superiority of Transformer models over RNNs; however, they need to specifically explore the integration of sentiment analysis within these frameworks for individual stock predictions. Furthermore, while Sidek et al. (2023) and Ridhawi and Osman (2023) explored sentiment analysis from news and social media, it could be beneficial for them to delve into the comparative analysis of different sentiment analysis tools or their combined impact on stock price predictions for specific companies.

This study addresses gaps in the existing literature by integrating sentiment scores from FinBERT, a financial sentiment analysis tool, and TextBlob, a general-purpose sentiment analysis tool, into Bidirectional Long Short-Term Memory (Bi-LSTM) networks and Multi-Layer Perceptrons (MLPs) to predict Nvidia's stock prices. By combining FinBERT’s nuanced financial sentiment insights with TextBlob’s broader analytical capabilities, this study provides a comprehensive framework for market sentiment analysis.

This approach extends beyond the scope of previous works, such as Cui et al. (2022) and Yekrangi and Nikolov (2023), by explicitly incorporating sentiment-driven insights into Bi-LSTM and MLP models for stock price prediction rather than focusing solely on text summarization and embedding techniques.

This study bridges the gap between generic sentiment analysis applications and targeted company-specific forecasting through a detailed case study of Nvidia Corporation. The proposed model evaluates the comparative predictive efficacy of FinBERT and TextBlob sentiment scores and demonstrates the potential benefits of integrating these tools into machine learning models. Furthermore, it examines the performance of Bi-LSTM and MLP architectures across diverse market conditions, highlighting the influence of sentiment integration on predictive accuracy.

This study theoretically and practically contributes to financial forecasting by offering insights into applying advanced natural language processing (NLP) tools and machine learning techniques. These findings can enhance predictive accuracy and inform decision-making in sentiment-driven financial modeling.

# 3. Methodology

## 3.1. Data collection and preprocessing

This section details the methodology employed in the comparative analysis of Bidirectional Long Short-Term Memory (Bi-LSTM) networks and Multi-Layer Perceptrons (MLPs) for predicting the stock prices of Nvidia Corporation. This study focuses on integrating market sentiments derived from financial news headlines and explores their impact on predictive accuracy across different market phases.

The study used two principal datasets spanning January 1, 2023, to March 31, 2024. The first dataset, obtained from Yahoo Finance, comprises 390 daily historical stock price records for Nvidia Corporation, encompassing metrics, such as opening, closing, highest, and lowest prices, and trading volume. The second dataset includes 9,400 news headlines about Nvidia during the same timeframe, measuring market sentiment toward the company. These data were obtained from Nasdaq.com.

Data preprocessing involved cleaning the datasets, including correcting anomalies, removing duplicates, handling missing or improperly formatted data, and splitting the data into training and testing sets. The train-test split was performed using an 80/20 ratio. Because the data is a time series dataset, shuffling was applied to ensure randomness while preserving the integrity of the temporal relationships within the dataset. This approach prevents the model from overfitting to specific patterns in sequential data while enabling a robust evaluation of its generalizability. Data cleaning is finding and rectifying (or removing) incorrect data, out-of-date, redundant, incomplete, or formatted incorrectly. Data cleaning was executed to correct anomalies and errors.

## 3.2. Feature engineering

In this study, we derived sentiment scores from financial news headlines to integrate market sentiments into Nvidia's stock price predictions. In addition, lags of the target variable were created to capture the temporal dependencies in stock price movements.

Sentiment analysis of the collected news headlines was conducted using two distinct approaches: TextBlob and FinBERT. FinBERT is a BERT-based natural language processing model optimized for financial sentiment analysis. It is pre-trained on financial corpora, such as Reuters TRC2, and fine-tuned on datasets like the Financial PhraseBank (Yang et al., 2020). By leveraging Hugging Face's pytorch_pretrained_bert, FinBERT employs advanced techniques, such as discriminative learning rates and gradual unfreezing, to prevent catastrophic forgetting during fine-tuning (Araci, 2019). These features allow FinBERT to deliver precise sentiment classification, making it a powerful tool for financial text analysis.

FinBERT, which is a fine-tuned BERT model, uses deep learning to analyze financial text, including financial jargon and sentiment precisely (Liu et al., 2023). Owing to its architecture, it excels at understanding finance-related topics. Consequently, conducting sentiment analysis on financial news and reports using FinBERT can be advantageous. It outperforms other methods when sentence structures become complex because BERT can encode a language in a universal representation (Gössi et al., 2023).

In contrast, TextBlob features a simple implementation for sentiment analysis, and its simplicity and flexibility make it applicable to all types of texts (Palomino et al., 2020). TextBlob is a Python library for natural language processing that leverages NLTK and Pattern to streamline core NLP tasks through an intuitive API. It processes text using TextBlob objects, which support operations such as part-of-speech tagging, noun phrase extraction, and sentiment analysis by employing statistical and rule-based algorithms (Loria, 2013). Sentiment analysis calculates polarity and subjectivity using pretrained models, whereas noun phrase extraction and tagging rely on Pattern's parsers. Spelling correction uses Norvig's algorithm, and lexical normalization features, such as lemmatization, are powered by WordNet. Extensible with custom tokenizer, taggers, and classifiers, TextBlob provides a flexible framework for developing efficient text analysis workflows.

Although TextBlob is not oriented toward financial texts, its general sentiment analysis capability can serve as a benchmark for primary sentiment measurement across various types of texts. Combining FinBERT and TextBlob could provide a comprehensive sentiment analysis tool by integrating FinBERT's capability of analyzing financial text with TextBlob's simplicity and widespread use (Hajek et al., 2022). The combined use of these tools allows for initial sentiment analysis using TextBlob, followed by the application of FinBERT procedures for deeper, finance-specific sentiment analysis, yielding optimal sentiment analysis results.

Sentiment scores for daily headlines were derived using FinBERT and TextBlob from a dataset of 9,400 Nvidia-related headlines and subsequently incorporated into Nvidia's historical stock data. TextBlob provides a general-purpose sentiment analysis framework, capturing broader sentiment patterns, while FinBERT offers domain-specific precision tailored to financial texts. The complementary nature of these models makes them ideal for joint application in financial forecasting. The scores were aggregated daily to align with the temporal resolution of the stock data and were subsequently employed as key variables in four distinct models. To ensure consistency, the sentiment scores were normalized and aggregated using a weighted average approach to account for varying headline counts per day. This integration ensured the creation of a comprehensive dataset that combined sentiment analysis and stock performance metrics, forming the modeling framework's foundation.

## 3.3. Modeling

This study investigates the predictive power of stock price forecasting models using two primary architectures: Bidirectional Long Short-Term Memory (Bi-LSTM) networks and Multi-Layer Perceptrons (MLPs). Each architecture is explored across two distinct temporal frameworks: the entire observation period and segmented market phases (Bullish Phase, Stagnation Phase, and Strong Bullish Phase). To ensure robust experimentation, four variations of the models were implemented, differing in their incorporation of sentiment analysis and feature engineering.

**Figure 1**

*Market Phases*

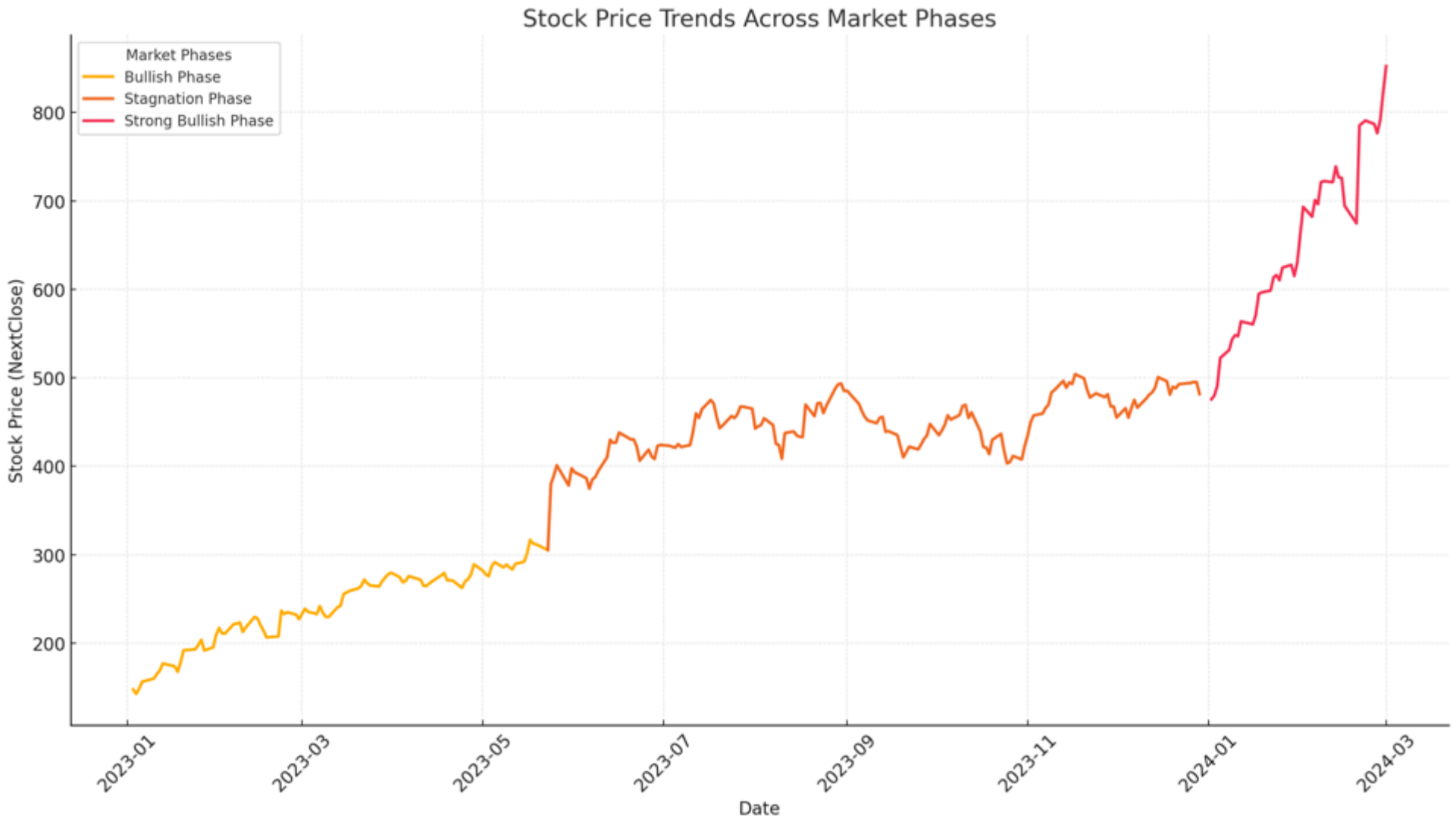

Bi-LSTM (Whole Period): Bidirectional LSTM models for the entire period are designed with layered configurations to capture sequential dependencies in stock price movements. Each model comprised two LSTM layers with 128 and 64 units, enhanced by dropout regularization (20%-30%) to mitigate overfitting. A dense output layer is employed for regression tasks, and the Adam optimizer (initial learning rate: 0.001) is used to update model weights. The early stopping with a patience of 10 epochs and adaptive learning rate adjustment via ReduceLROnPlateau ensured efficient training. The input features include scaled historical prices (Daily Close) and sentiment scores (FinBERT, TextBlob, or both), depending on the model configuration. A sliding window mechanism was applied to structure the input data for sequential learning.

Bi-LSTM (Market Phases): Bi-LSTM models are independently trained on data from each defined phase to investigate the impact of different market conditions. The architecture is similar to the whole-period Bi-LSTM but is tailored to phase-specific data. Distinctive hyperparameter adjustments, such as dropout rates (0.25–0.3) and a slightly higher initial learning rate (0.003), were made to accommodate the phase-specific characteristics. Separate models were trained for the Bullish, Stagnation, and Strong Bullish phases. This segmentation allows us to analyze whether market conditions influence predictive performance.

MLP (Whole Period): The MLP models are fully connected architectures featuring three dense layers (256, 128, and 64 units) and dropout layers (30%-40%) for regularization. Batch normalization is implemented for training stability, and MinMaxScaler is applied to normalize the input features. In addition to historical prices, lagged features such as Daily_Close_Lag1, Lag2, and Lag3 are generated to capture temporal dependencies. The model was optimized using Adam or RMSprop with a tuned learning rate (0.003–0.01) and regularization (L2 penalty of 0.001). The architecture's simplicity contrasts with the sequential nature of Bi-LSTM, allowing for a direct comparison of performance under varying feature sets.

MLP (Market Phases): For the segmented market phases, the MLP models are restructured with two dense layers (128 and 64 units) to align with the reduced data volume. Moving averages (MA_5, MA_10) and volatility measures are added as input features to complement historical prices and sentiment scores. The dropout rate was fixed at 30%, and additional fine-tuning of the dense layer units was performed to avoid underfitting. Each market phase was treated independently, and the models were trained on phase-specific data splits. The learning rates are dynamically adjusted using ReduceLROnPlateau, thereby enabling efficient convergence.

In all models, the input features were systematically varied to evaluate the impact of sentiment analysis on stock price prediction. The baseline model (Model 1) exclusively uses historical daily closing prices (Daily Close) as a control. Models 2 and 3 integrate the sentiment scores derived from FinBERT and TextBlob, respectively, to provide a comparative view of nuanced and general sentiment impacts. Model 4 combines sentiment sources with historical data to determine whether integrating diverse sentiment analysis techniques enhances predictive accuracy.

Hyperparameter optimization is performed systematically for each architecture, focusing on tuning the number of LSTM units, dropout rates, and batch sizes. At the same time, MLPs emphasize dense layer configurations, learning rates, and regularization techniques. Both architectures employ early stopping and adaptive learning rate schedules to prevent overfitting and ensure optimal performance. The influence of these hyperparameters was evaluated by validating the performance of the proposed models and temporal frameworks.

## 3.4. Results

Evaluating the performance of quantitative models often relies on statistical metrics such as the mean absolute error (MAE), Mean Squared Error (MSE), Root Mean Squared Error (RMSE), and coefficient of determination ($R^2$). These metrics offer distinct insights into model accuracy and error distribution, which are context-dependent.

The MAE represents the average magnitude of errors, providing a straightforward measure of prediction accuracy without considering error directionality and is less sensitive to variability than the MSE or RMSE (Willmott and Matsuura, 2005). The MSE, by squaring errors, amplifies larger deviations and allows decom-position into systematic and unsystematic components, offering a deeper understanding of prediction errors (Robeson and Willmott, 2023).

RMSE, as the square root of MSE, is interpretable in the same units as the data and is suitable for Gaussian error distributions, making it appropriate for assessing model performance under these conditions (Chai and Draxler, 2014). RMSE satisfies the triangle inequality requirement for a distance metric, reinforcing its mathematical validity, but its sensitivity to error variability can make it misleading for summarizing average errors (Willmott et al., 2009).

$R^2$ quantifies the proportion of variance in the dependent variable explained by the model, offering a normalized measure of accuracy and making it particularly useful for assessing the overall predictive power of regression models (Chicco et al., 2021).

There is no universally superior metric; the choice between MAE, MSE, RMSE, and $R^2$ depends on the application context and error characteristics. RMSE may provide more reliable insights for Gaussian error distributions, whereas MAE offers a robust measure under non-Gaussian conditions (Hodson, 2022). $R^2$ complements these metrics by providing a variance-based perspective, enabling researchers to capture overall model explanatory power (Chicco et al., 2021).

This study employed feature importance analysis via ablation feature analysis. In addition to statistical metrics, ablation-based techniques provide valuable insights into model performance by systematically analyzing the contributions of individual features. This approach offers a rigorous evaluation of feature importance, particularly in improving model precision and predictability, as it systematically removes features to observe their impact on predictions (Pali & Tiwari, 2024). Furthermore, ablation methods focus on analyzing variations in the input space, offering a computationally efficient alternative to traditional output space-based approaches (Yang et al., 2023).

3.4.1. Bi-LSTM Model for the Whole Period

Figure 2 shows that the Bi-LSTM models generally align more closely with actual values than MLP models, especially in advanced versions that incorporate text sentiment analysis with FinBERT and TextBlob. However, some over-prediction is evident, particularly at peaks.

**Figure 2**

*Bi-LSTM Architecture: Actual vs. Predicted for the Whole Period*

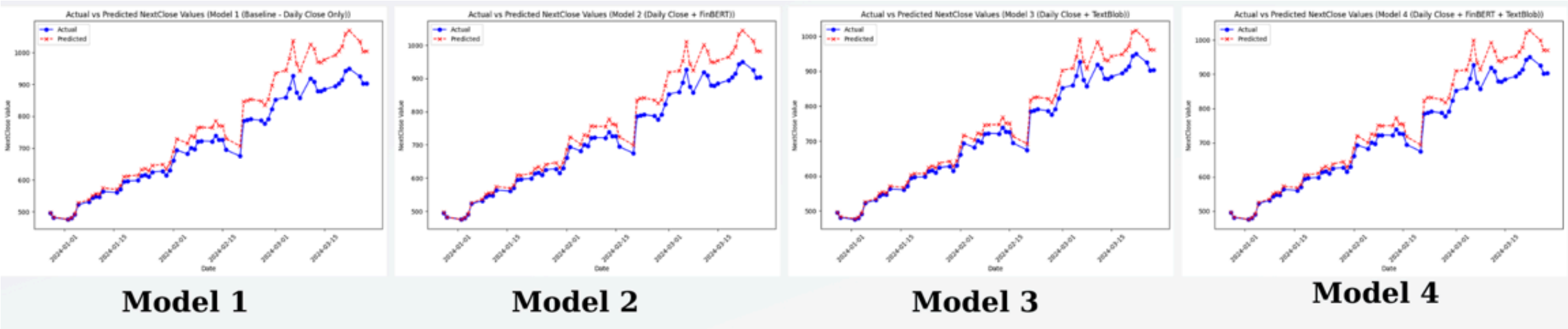

Table 1 demonstrates that integrating sentiment scores into the Bi-LSTM models enhances predictive performance compared to the baseline model using only historical closing prices. For clarity, the models in the table are referenced as Model 1 through Model 4, corresponding to the following configurations: (1) Baseline with daily close prices only, (2) FinBERT sentiment scores, (3) TextBlob sentiment scores, and (4) Combined sentiment features. TextBlob scores led to the lowest MAE, MSE, and RMSE, along with a higher $R^2$ Score. The combined sentiment model (FinBERT + TextBlob) also improved over the baseline model; however, the effect was less pronounced than with TextBlob alone. The substantial discrepancy between the train and test MAE (e.g., Model 3: 1.60 vs. 30.19) raises concerns about potential overfitting; however, the maintained high test $R^2$ (0.93) suggests that while generalization may be limited, the model retains explanatory power on unseen data.

**Table 1**

*Performance of Bi-LSTM over the Whole Period*

| Metric | Model 1 | Model 2 | Model 3 | Model 4 |
|---|---|---|---|---|
| Train MAE | 2.3559 | 2.7344 | 1.6021 | 2.5229 |
| Train MSE | 9.9913 | 10.5720 | 4.5054 | 9.6811 |
| Train RMSE | 3.1609 | 3.2515 | 2.1226 | 3.1114 |
| Train $R^2$ | 0.9991 | 0.9991 | 0.9996 | 0.9991 |
| Test MAE | 48.2915 | 37.5577 | 30.1905 | 30.5288 |
| Test MSE | 3620.4197 | 2166.1026 | 1384.7762 | 1471.3825 |
| Test RMSE | 60.1699 | 46.5414 | 37.2126 | 38.3586 |
| Test $R^2$ | 0.8266 | 0.8963 | 0.9337 | 0.9295 |

The MAE and RMSE provide direct measures of prediction error; thus, reducing these metrics suggests improving model accuracy (Willmott & Matsuura, 2005). However, interpreting the high $R^2$ Score requires caution because high values may not always indicate better predictive models in complex datasets (Kovacs et al., 2008).

**Table 2**

*Bi-LSTM Feature Importance Over the Whole Life Cycle*

| Feature | Model 1 | Model 2 | Model 3 | Model 4 |
|---|---|---|---|---|
| Daily Close | 599.38446 | 605.831726 | 595.140259 | 600.952454 |
| FinBERT | - | 20.701258 | - | 19.360909 |
| TextBlob | - | - | 6.573365 | 3.991386 |

Table 2 presents the Bi-LSTM feature importance over the entire period. The Daily Close feature dominates across all models, with minimal variation in magnitude (e.g., 599.38–605.83), suggesting its consistent predictive utility. FinBERT exhibits moderate importance in Models 2 and 4 (20.70 and 19.36, respectively), while TextBlob shows limited influence (6.57 in Model 3, declining to 3.99 in Model 4). This may indicate that domain-specific sentiment (FinBERT) contributes more robustly than general sentiment (TextBlob) when combined with price data.

### 3.4.2. Bi-LSTM Model for the Bullish Phase

Figure 3 illustrates that the Bİ-LSTM models capture the upward trend more effectively in the bullish phase than in the other phases. However, some models exhibit a lag in their prediction responsiveness to steep increases. Advanced models integrating FinBERT and sentiment analysis showed improvements in tracking these trends but still underestimated sharp bullish movements.

**Figure 3**

*Bi-LSTM Architecture: Actual vs. Predicted for Market Phases*

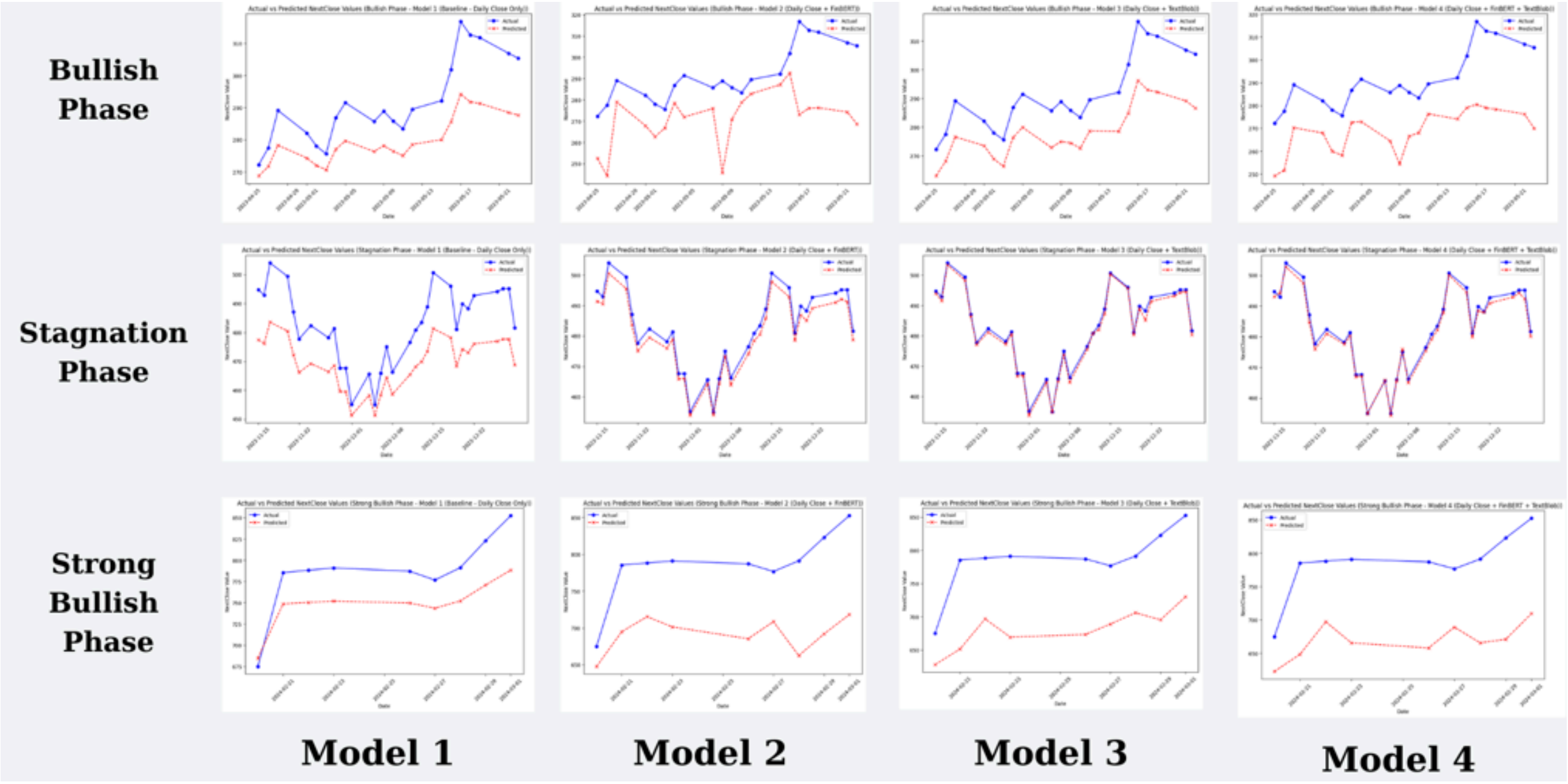

Table 3 reveals that the baseline model yielded lower errors in the Bullish Phase than the sentiment-based models. This suggests that the sentiment features have limited predictive value when capturing the dynamics of bullish markets in this dataset. The scikit-learn documentation notes that $R^2$ is not symmetric and may yield negative values, indicating poor model performance (www.scikit-learn.org, 2024). This underscores the deficiencies of the models in capturing the underlying data patterns, reflecting the inadequate regression performance (Chicco et al., 2021). In this case, the negative test $R^2$ values across models (e.g., Model 4: −3.69) indicate poor generalization during this phase and complicate straightforward overfitting diagnosis.

**Table 3**

*Bİ-LSTM Model Performance During the Bullish Phase*

| Metric | Model 1 | Model 2 | Model 3 | Model 4 |
|---|---|---|---|---|
| Train MAE | 17.7786 | 25.9949 | 18.4805 | 27.0010 |
| Train RMSE | 23.7529 | 33.2531 | 23.8700 | 34.4501 |
| Train $R^2$ | 0.6152 | 0.2458 | 0.6114 | 0.1905 |
| Test MAE | 12.2221 | 19.3260 | 12.6301 | 23.6121 |
| Test RMSE | 13.4296 | 22.9737 | 14.6306 | 27.7661 |
| Test $R^2$ | −0.0962 | -2.2081 | -0.3011 | -3.6861 |

Table 4 examines the Bi-LSTM model during the bullish phase. The Daily Close importance declines substantially in Models 2 and 4 (53.67 and 48.03) relative to Model 1 (82.23), coinciding with FinBERT's heightened importance (43.21–43.35). TextBlob remains comparatively marginal (9.45–12.77), implying that bullish phases may amplify the relevance of financial sentiment analysis while tempering reliance on raw price data. These results suggest contextual interdependence between market conditions and feature utility.

**Table 4**

*Bİ-LSTM Model's Feature Importance during the Bullish Phase*

| Feature | Model 1 | Model 2 | Model 3 | Model 4 |
|---|---|---|---|---|
| Daily Close | 82.22802 | 53.671803 | 77.847878 | 48.030838 |
| FinBERT | - | 43.208466 | - | 43.347618 |
| TextBlob | - | - | 12.769716 | 9.446444 |

### 3.4.3. Bi-LSTM Model for Stagnation Phase

Figure 3 reveals that during stagnation, the Bi-LSTM models display relatively close alignment between the actual and predicted values, particularly in advanced models. However, basic models need help handling volatility and tend to over-smooth fluctuations, leading to discrepancies during local peaks and troughs.

**Table 5**

*Bİ-LSTM Model Performance During the Stagnation Phase*

| Metric | Model 1 | Model 2 | Model 3 | Model 4 |
|---|---|---|---|---|
| Train MAE | 8.0616 | 2.9153 | 1.3732 | 1.9244 |
| Train RMSE | 10.4995 | 4.0719 | 2.1984 | 2.7464 |
| Train $R^2$ | 0.8771 | 0.9815 | 0.9946 | 0.9916 |
| Test MAE | 13.3956 | 3.6290 | 1.1143 | 1.1935 |
| Test RMSE | 13.9982 | 3.8472 | 1.1613 | 1.5051 |
| Test $R^2$ | −0.1791 | 0.9109 | 0.9919 | 0.9864 |

Table 5 indicates that models that incorporate sentiment features performed significantly better during the Stagnation Phase than the baseline. A close alignment between train and test metrics (e.g., Model 3 MAE: 1.37 vs. 1.11; $R^2$: 0.99 vs. 0.99) suggests minimal overfitting risks, with sentiment-enhanced models demonstrating robust generalization. This may be attributed to TextBlob's ability to capture generalized sentiment patterns, which are more impactful during less volatile market conditions where macroeconomic or external sentiment dominates. The model using TextBlob scores achieved the lowest errors and the highest $R^2$ Score, which, when considered alongside the reduced MAE and RMSE, suggested improved predictive capability (Chai & Draxler, 2014). However, as with previous phases, the high $R^2$ value should be interpreted cautiously (Kovacs et al., 2008).

Table 6 presents the evaluation of the stagnation phase of Bi-LSTM. Daily Close importance increased across all models (114.65–160.02), whereas FinBERT (3.22–3.26) and TextBlob (0.44–1.71) exhibited negligible values. This implies that technical factors may supersede sentiment analysis in low-volatility environments, potentially due to reduced market-moving news or sentiment-driven trading activity during stagnation periods.

**Table 6**

*Bİ-LSTM Model's Feature Importance during the Stagnation Phase*

| Feature | Model 1 | Model 2 | Model 3 | Model 4 |
|---|---|---|---|---|
| Daily Close | 114.646614 | 160.022354 | 156.829514 | 155.978226 |
| FinBERT | - | 3.256613 | - | 3.218168 |
| TextBlob | - | - | 1.712705 | 0.435842 |

### 3.4.4. Bi-LSTM: Strong Bullish Phase

Figure 3 highlights that Bİ-LSTM models in the strong bullish phase demonstrate clear limitations in accurately tracking steep upward trends, with a significant underprediction bias in primary and even advanced models. The baseline model yields the best results; however, those incorporating richer features (e.g., sentiment) demonstrate incremental improvement.

**Table 7**

*Bİ-LSTM Model Performance During the Strong Bullish Phase*

| Metric | Model 1 | Model 2 | Model 3 | Model 4 |
|---|---|---|---|---|
| Train MAE | 39.8590 | 53.8672 | 55.4161 | 58.8593 |
| Train RMSE | 49.8389 | 65.1150 | 68.5951 | 69.8090 |
| Train $R^2$ | 0.5872 | 0.2953 | 0.2180 | 0.1901 |
| Test MAE | 40.6381 | 79.8523 | 108.8950 | 107.6490 |
| Test RMSE | 43.1470 | 85.8766 | 112.2521 | 111.6134 |
| Test $R^2$ | 0.0836 | −2.6302 | −5.2026 | −5.1322 |

Table 7 demonstrates that in the Strong Bullish Phase, all sentiment-based models exhibited higher errors and lower predictive performance than the baseline. The negative $R^2$ Scores indicate that the models perform poorly relative to the mean model (Chicco et al., 2021). Negative $R^2$ scores occur when the model's predictions are less accurate than simply using the mean of the target variable, signifying poor alignment with the data's underlying patterns. This suggests that sentiment scores may not align well with market behavior under strong bullish trends in this context. The increased MAE and RMSE values further support these observations. After analyzing the performance of the Bi-LSTM models, we now turn to MLP architectures to explore how simpler models perform under similar conditions.

Table 8 presents the strong bullish phase of Bi-LSTM. Daily Close importance declined sharply in Models 2–4 (59.74–92.12 vs. 156.08 in Model 1), while FinBERT (55.21–67.81) and TextBlob (22.48–31.08) gained prominence. Phase-specific amplification of sentiment features suggests heightened investor sensitivity to sentiment signals during pronounced upward trends, possibly reflecting momentum-driven trading behavior.

**Table 8**

*Bi-LSTM Model's Feature Importance during the Strong Bullish Phase*

| Feature | Model 1 | Model 2 | Model 3 | Model 4 |
|---|---|---|---|---|
| Daily Close | 156.07869 | 92.123726 | 110.067162 | 59.740452 |
| FinBERT | - | 67.810661 | - | 55.208836 |
| TextBlob | - | - | 31.075907 | 22.479275 |

### 3.4.5. MLP Model for the Whole Period

MLP models were evaluated to assess their ability to predict stock prices across the entire observation period. Four variations were tested, including baseline historical price inputs and models augmented with sentiment scores from FinBERT, TextBlob, and both. The evaluation focused on understanding how well MLP models captured temporal dependencies and incorporated sentiment-based features to enhance predictive accuracy.

**Figure 4**

*MLP Architecture: Actual vs. Predicted for the Whole Period*

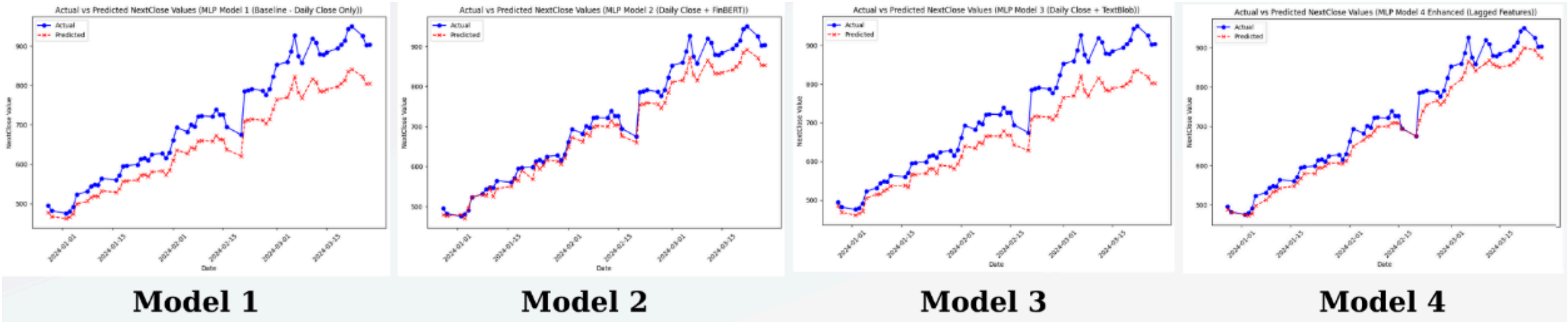

Figure 3 shows that the MLP models generally underperform compared to Bİ-LSTM models in terms of overall alignment with the actual values. Basic MLP models, in particular, fail to capture trend dynamics accurately and often show lower amplitudes in predictions than in actual fluctuations.

**Table 9**

*MLP Model Performance Over the Whole Period*

| Metric | Model 1 | Model 2 | Model 3 | Model 4 |
|---|---|---|---|---|
| Train MAE | 7.9783 | 6.4946 | 4.1648 | 6.5516 |
| Train MSE | 79.9371 | 84.4271 | 27.9430 | 71.5400 |
| Train RMSE | 8.9408 | 9.1884 | 5.2861 | 8.4581 |
| Train $R^2$ | 0.9927 | 0.9922 | 0.9974 | 0.9934 |
| Test MAE | 63.0017 | 32.4626 | 47.7904 | 26.4368 |
| Test MSE | 4938.6566 | 1345.1944 | 3648.4341 | 941.7557 |
| Test RMSE | 70.2756 | 36.6769 | 60.4023 | 30.6880 |
| Test $R^2$ | 0.7635 | 0.9356 | 0.8253 | 0.9549 |

Table 9 shows that integrating FinBERT sentiment scores into the MLP model significantly improved predictive performance. The reductions in MAE, MSE, and RMSE indicate better accuracy (Willmott & Matsuura, 2005), while the high $R^2$ Score suggests a good fit. However, as previously noted, $R^2$ should be interpreted cautiously and in conjunction with other error metrics (Kovacs et al., 2008). In addition, model 4 with FinBERT and TextBlob demonstrated the lowest errors, which indicates that these additional inputs enhance the performance in this case. Model 4's elevated test MAE (26.44) relative to training (6.55) signals potential overfitting, but its high test $R^2$ (0.95) tempers this interpretation, instead implying constrained but functional generalization capacity.

Table 10 outlines MLP feature importance over the full period. Model 4 introduces lagged Daily Close variables (Lag1–Lag3: 112.77–116.23), which collectively surpass the contemporaneous Daily Close (140.09). FinBERT (16.13) and TextBlob (8.91) exhibit reduced importance compared to Bi-LSTM results in Table 2, highlighting architectural differences in feature processing. The prominence of lagged features in MLP may reflect its capacity to leverage sequential dependencies that are less accessible to Bi-LSTM.

**Table 10**

*MLP Feature Importance Over the Whole Period*

| Feature | Model 1 | Model 2 | Model 3 | Model 4 |
|---|---|---|---|---|
| Daily Close | 559.399048 | 468.888977 | 515.688171 | 140.091049 |
| FinBERT | - | 39.443817 | - | 16.132677 |
| TextBlob | - | - | 7.249935 | 8.909840 |

| Feature | Model 1 | Model 2 | Model 3 | Model 4 |
|---|---|---|---|---|
| Daily Close_Lag1 | - | - | - | 112.771538 |
| Daily Close_Lag2 | - | - | - | 116.225800 |
| Daily Close_Lag3 | - | - | - | 82.880028 |

### 3.4.6. MLP Model-Bullish Phase

Figure 5 demonstrates that MLP models struggle during bullish, with significant under-prediction biases, especially for sharp upward movements. Even enhanced models with sentiment features must close this gap effectively.

**Figure 5**

*MLP Architecture: Actual vs. Predicted for Market Phases*

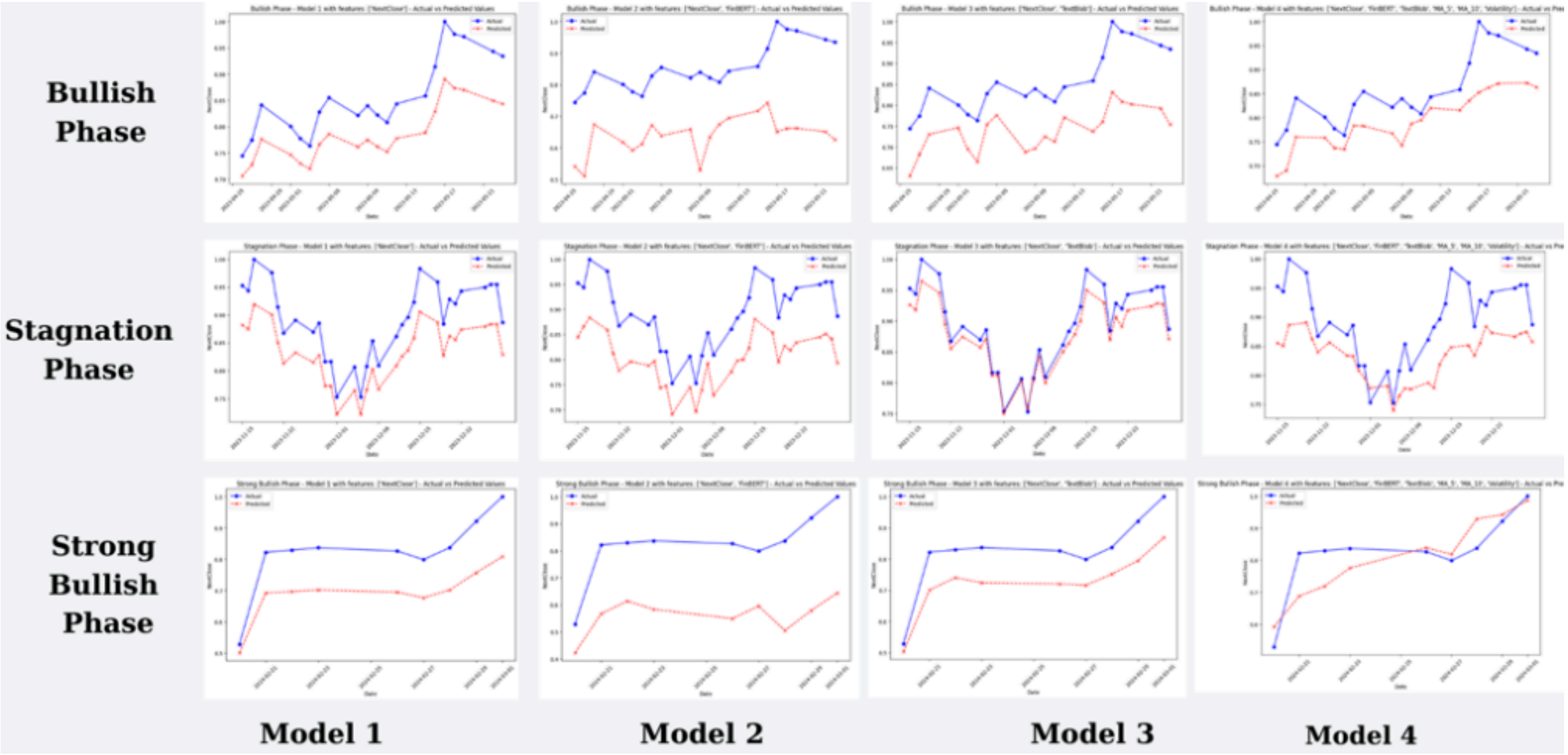

Figure 5 highlights the performance of MLP models across market phases, demonstrating that incorporating sentiment features improves predictive accuracy in certain contexts.

**Table 11**

*MLP Model Performance During the Bullish Phase*

| Metric | Model 1 | Model 2 | Model 3 | Model 4 |
|---|---|---|---|---|
| Train MAE | 0.0511 | 0.1020 | 0.0536 | 0.0347 |
| Train RMSE | 0.0634 | 0.1260 | 0.0705 | 0.0390 |
| Train $R^2$ | 0.9168 | 0.6717 | 0.8973 | 0.9685 |
| Test MAE | 0.0701 | 0.1514 | 0.0721 | 0.0669 |
| Test RMSE | 0.0727 | 0.1627 | 0.0755 | 0.0698 |
| Test $R^2$ | 0.0258 | -3.8800 | −0.0513 | 0.1030 |

Table 11 indicates that the baseline model demonstrated comparable or slightly better performance across metrics in the Bullish Phase. The inclusion of sentiment features did not enhance predictive accuracy and, in some cases, increased the number of errors. Low and negative $R^2$ scores highlight fundamental model mismatch with bullish dynamics, prioritizing poor generalization over overfitting.

Table 12 analyzes MLP during bullish phases. Model 4 shows significant contributions from MA_5 (42.23) and MA_10 (52.64), overshadowing contemporaneous Daily Close (32.56) and sentiment features (FinBERT: 1.16; TextBlob: 2.37). This contrasts with Bi-LSTM results (Table 4), in which sentiment played a larger role, suggesting MLP's greater reliance on technical indicators during bullish conditions.

**Table 12**

*MLP Model Feature Importance During the Bullish Phase*

| Feature | Model 1 | Model 2 | Model 3 | Model 4 |
|---|---|---|---|---|
| Daily Close | 119.399246 | 49.058277 | 95.974411 | 32.558304 |
| FinBERT | - | 39.111164 | - | 1.162262 |
| TextBlob | - | - | 5.202605 | 2.371144 |
| MA_5 | - | - | - | 42.227772 |
| MA_10 | - | - | - | 52.640247 |
| Volatility | - | - | - | 2.467628 |

### 3.4.7. MLP Model for the Stagnation Phase

Figure 4-2 indicates that, in the stagnation phase, MLP models show some improvement in capturing the trends compared to the general Bİ-LSTM models. Incorporating only TextBlob sentiment yields the best results.

**Table 13**

*MLP Model Performance During the Stagnation Phase*

| Metric | Model 1 | Model 2 | Model 3 | Model 4 |
|---|---|---|---|---|
| Train MAE | 0.0267 | 0.0324 | 0.0170 | 0.0623 |
| Train RMSE | 0.0348 | 0.0423 | 0.0253 | 0.0762 |
| Train $R^2$ | 0.9467 | 0.9214 | 0.9717 | 0.7441 |
| Test MAE | 0.0589 | 0.0730 | 0.0224 | 0.0986 |
| Test RMSE | 0.0604 | 0.0756 | 0.0243 | 0.1053 |
| Test $R^2$ | 0.1333 | −0.3596 | 0.8592 | −1.6347 |

Table 13 shows that the model that incorporates TextBlob sentiment scores exhibits superior predictive performance. Model 3's strong test performance (MAE: 0.02 vs. train 0.02; $R^2$: 0.86 vs. 0.97) indicates effective generalization during stagnation, with no conclusive evidence of overfitting despite slight $R^2$ attenuation. This observation aligns with the Bİ-LSTM models, in which sentiment features strongly influence price behavior during stagnant conditions.

Table 14 assesses MLP during stagnation. MA_5 (51.34) remains influential in Model 4, whereas MA_10 (10.57) and the sentiment features exhibit limited importance. Daily Close retains moderate predictive value (55.96), which is consistent with Bi-LSTM findings (Table 6) in underscoring the dominance of technical factors during stagnant phases, albeit with MLP prioritizing moving averages over raw price data.

**Table 14**

*MLP Model Feature Importance During the Stagnation Phase*

| Feature | Model 1 | Model 2 | Model 3 | Model 4 |
|---|---|---|---|---|
| Daily Close | 139.35379 | 124.340477 | 135.508286 | 55.959759 |
| FinBERT | - | 3.648912 | - | 1.482877 |

| Feature | Model 1 | Model 2 | Model 3 | Model 4 |
|---|---|---|---|---|
| TextBlob | - | - | 1.479448 | 2.298968 |
| MA_5 | - | - | - | 51.340214 |
| MA_10 | - | - | - | 10.568443 |
| Volatility | - | - | - | 0.626636 |

### 3.4.8. MLP Model-Strong Bullish Phase

Figure 4-3 reveals that MLP models exhibit substantial underprediction during the strong bullish phase. The predictions do not capture the sharp increase in actual values, even in models with enhanced features. This highlights the model's limitations when considering high volatility or strongly directional phases.

**Table 15**

*MLP Model Performance During the Strong Bullish Phase*

| Metric | Model 1 | Model 2 | Model 3 | Model 4 |
|---|---|---|---|---|
| Train MAE | 0.0608 | 0.1369 | 0.0301 | 0.0613 |
| Train RMSE | 0.0706 | 0.1591 | 0.0368 | 0.0777 |
| Train $R^2$ | 0.8824 | 0.4028 | 0.9681 | 0.8577 |
| Test MAE | 0.1152 | 0.2798 | 0.0947 | 0.0669 |
| Test RMSE | 0.1214 | 0.2925 | 0.0978 | 0.0731 |
| Test $R^2$ | −0.0299 | −4.9775 | 0.3320 | 0.6268 |

Table 15 demonstrates that Model 4, which integrates FinBERT and TextBlob sentiment features with daily close prices, demonstrated the best performance. Model 4's moderate train-test divergence (MAE: 0.06 vs. 0.07; $R^2$: 0.86 vs. 0.63) suggests limited overfitting, although reduced test $R^2$ reflects challenges in extrapolating to extreme bullish conditions. This indicates that the combination of sentiment features enhances the model's ability to capture meaningful patterns in the data.

Table 16 evaluates MLP during strong bullish phases. Model 4 highlights MA_5 (73.53) and MA_10 (55.13) as primary drivers, with TextBlob importance (27.16) being elevated compared to the other phases. FinBERT's contribution rose to 11.57, contrasting with its minimal role in Table 12. This phase-specific resurgence in sentiment features in MLP, paired with strong moving average signals, may reflect the compound effects of technical and sentiment factors during extreme market movements.

**Table 16**

*MLP Model Feature Importance During the Strong Bullish Phase*

| Feature | Model 1 | Model 2 | Model 3 | Model 4 |
|---|---|---|---|---|
| Daily Close | 232.318054 | 254.964844 | 248.268127 | 34.119873 |
| FinBERT | - | 5.209588 | - | 11.571093 |
| TextBlob | - | - | 21.933254 | 27.159782 |
| MA_5 | - | - | - | 73.530518 |
| MA_10 | - | - | - | 55.126667 |
| Volatility | - | - | - | 6.010790 |

The predictive performance of the models varied across different architectures, feature sets, and market phases. Sentiment-based features, particularly TextBlob scores, showed substantial improvements in specific scenarios, such as the Stagnation Phase. However, their effectiveness appeared to diminish under strongly bullish conditions. The impact of integrating FinBERT and lagged features also varied depending on the temporal framework and model architecture. These findings suggest that the utility of sentiment features may depend on specific market dynamics, warranting further investigation into feature selection and market-specific modeling approaches.

## 4. Discussion

The integration of sentiment analysis and machine learning models to predict stock prices has been a significant area of research in financial studies. This study explores the integration of two sentiment analysis tools—FinBERT and TextBlob—into predictive models focused on Nvidia Corporation's stock price trends. By evaluating the effectiveness of these tools across distinct market phases and comparing their outputs, this research provides insights into the dynamics of market sentiment and its influence on stock price prediction. This section discusses the study's contributions, its relationship to previous works, and its implications for the existing literature.

The combined application of FinBERT, a financial-domain-specific sentiment analysis tool, and TextBlob, a general-purpose sentiment analysis tool, represents an approach to capture different facets of market sentiment from financial news headlines. FinBERT excels in parsing and interpreting complex financial reports, such as earnings reports and market analyses, offering nuanced sentiment scores tailored to the finance domain. TextBlob, on the other hand, provides a more generalized sentiment analysis framework, allowing for an alternative perspective on the same textual data. One reason for the differential performance across market phases is that domain-specific sentiment tools like FinBERT may underemphasize large-scale sentiment shifts, which often dominate strong bull markets and overshadow nuanced financial signals. In contrast, broad-spectrum tools such as TextBlob can capture more public sentiment; thus, they perform better in moderately moving or stagnant conditions, where external sentiment exerts a stronger influence.

Integrating these tools, this study captures a dual perspective on sentiment from the same news headlines. FinBERT delivers domain-specific precision by focusing on financial terminology and context, while TextBlob provides a general sentiment assessment that can capture broader linguistic cues. The proposed framework enhances sentiment analysis by leveraging specialized financial sentiment and general sentiment interpretation from the same textual data. This dual sentiment extraction provides a more comprehensive understanding of investor sentiment reflected in financial news headlines. Additionally, the feature importance findings reveal that "Daily Close" consistently remains the single most influential predictor, especially when market momentum is strong. In contrast, sentiment features gained prominence during low-volatility or stagnation phases, indicating the context-dependent value of both specialized (FinBERT) and general-purpose (TextBlob) sentiment tools.

This methodological approach aims to contribute to the literature by illustrating the potential complementary strengths of these tools and by offering empirical evidence of their impact on predictive accuracy across different market phases.

Hajek et al. (2022) explored the integration of sentiment analysis into financial forecasting and highlighted the effectiveness of combining sentiment features. However, their work primarily relied on generalized sentiment analysis techniques, limiting its applicability to specialized financial contexts. This study builds upon the work of Hajek et al. by incorporating FinBERT, a tool specifically pretrained on financial texts, with the aim of addressing the limitations of generalized approaches to financial forecasting.

Moreover, while Hajek et al. emphasized sentiment extraction as a standalone feature, this study takes a step further by integrating both FinBERT and TextBlob sentiment scores with temporal modeling techniques (Bi-LSTM and MLP) to capture both qualitative and quantitative drivers of stock price movements. By segmenting the analysis across distinct market phases (Bullish, Stagnation, and Strong Bullish), this study expands upon previous approaches, offering phase-specific insights into the role of sentiment.

This study aims to contribute to the literature by comparing sentiment analysis outputs from two different tools applied to the same financial news headlines, offering a perspective on their effectiveness. Unlike studies such as Cui et al. (2022), which focused on optimizing text summarization and embedding techniques without tailoring sentiment analysis to specific financial contexts, this study demonstrated the value of combining domain-specific sentiment tools with general-purpose ones for nuanced prediction.

Another significant contribution is the phase-specific analysis of sentiment's impact. The results show that sentiment analysis is particularly effective during stagnation phases, but its utility diminishes significantly in directional markets. These findings suggest that the interplay between sentiment and market behavior is dynamic and context-dependent, offering actionable insights for practitioners and researchers.

This study illustrates the potential benefits of combining different analytical perspectives from the same data source by applying FinBERT and TextBlob to financial news headlines and integrating their sentiment scores into predictive models. The proposed method captures different layers of sentiment interpretation and provides a rich dataset for predictive modeling.

This study addresses several critical aspects of the existing literature:

1. Comparative Analysis of Sentiment Tools: Previous studies have often focused on financial-domain-specific tools such as FinBERT or general-purpose tools like TextBlob, but rarely on the same dataset. By systematically comparing and integrating their outputs, this study comprehensively evaluates their relative strengths and weaknesses using financial news headlines, particularly in stock price prediction.
2. Phase-Specific Insights: Most previous studies evaluated sentiment analysis tools in aggregate without considering how their effectiveness varies across market conditions. This study’s segmentation of market phases (Bullish, Stagnation, and Strong Bullish) reveals the nuanced role of sentiment in different contexts, offering a more granular understanding of its predictive utility.
3. Integration of Multiple Sentiment Perspectives: This study introduces a dual-perspective sentiment framework by combining sentiment scores from FinBERT and TextBlob applied to the same financial news headlines. This approach captures domain-specific and general sentiment interpretations, thus addressing the limitations of relying on a single sentiment analysis tool.
4. Application to Firm-Specific Prediction: While many studies have adopted a market-level perspective, this study focuses on Nvidia Corporation and provides targeted insights that bridge the gap between market-level analysis and firm-specific forecasting. This specificity contributes to the growing body of research on how sentiment analysis can be tailored to individual companies.

This study contributes to the field by enhancing the integration of sentiment analysis into machine learning models for stock price prediction. The combination of FinBERT and TextBlob, which are applied to financial news headlines, phase-specific analysis, and firm-focused applications, fills essential areas in the literature. Although the findings suggest significant potential for improving predictive accuracy, they underscore the need for context-aware modeling and carefully selecting sentiment tools based on market conditions. These contributions offer valuable directions for future research, emphasizing the dynamic interplay between sentiment, machine learning, and financial forecasting.

## 5. Conclusion

This study explored the integration of sentiment analysis into stock price prediction models, focusing on Nvidia Corporation, and evaluated the performance of Bi-LSTM and MLP architectures. TextBlob sentiment scores consistently outperformed FinBERT, particularly during the Stagnation Phase, suggesting that generalized sentiment tools may better capture broader market psychology in specific contexts. Feature importance analysis revealed that “Daily Close” remained the dominant predictor across all market conditions, particularly during strong directional trends. However, sentiment features such as TextBlob gained higher relevance during stagnation phases, which emphasizes the need for context-aware modeling to maximize predictive accuracy. FinBERT effectively processed financial-specific text but had a limited impact during phases in which external market trends or general sentiment dominated. Combining FinBERT and TextBlob sentiment scores provided incremental benefits in specific cases, such as the Strong Bullish Phase for MLP models, but often showed redundancy.

In this study, Bi-LSTM models generally outperformed MLPs in terms of capturing temporal dependencies and sequential patterns, particularly during the Stagnation Phase and the overall period. However, MLPs demonstrated competitive performance during the Stagnation Phase, indicating their suitability for contexts in which sentiment plays a more significant role and simpler architectures are sufficient. Market phase sensitivity appears to be a critical factor, with sentiment features showing potential utility during stagnant markets but potentially reduced relevance during bullish phases dominated by technical trends.

The study's contributions involve the integration of FinBERT and TextBlob, offering a dual-layer sentiment perspective, and conducting a phase-specific analysis of sentiment's role in stock price prediction. By comparing Bi-LSTM and MLP architectures across market conditions, this study provides actionable insights into model selection and sentiment utilization, addressing gaps in the literature and advancing sentiment-driven financial forecasting.

Despite its contributions, this study has limitations. The analysis focused on Nvidia Corporation over a short timeframe, thereby limiting generalizability. Sentiment analysis relied on news headlines, which may only partially capture real-time market sentiment from broader sources like social media. In addition, the segmentation of market phases may need to be more balanced with the complexity of real-world market dynamics.

Future research could expand the dataset to include multiple companies and diverse sentiment sources, such as social media platforms. Exploring hybrid architectures, such as combining Bi-LSTM with transformers and developing models that dynamically adapt to changing market phases, could improve prediction accuracy. Overall, this study highlights the potential of sentiment analysis and context-aware modeling in financial forecasting and proposes directions for future research in this field.

Peer Review Externally peer-reviewed.
Conflict of Interest The author have no conflict of interest to declare.
Grant Support The author declared that this study has received no financial support.

Author Details **Ahmet Akusta (Lecturer Dr.)**
[1] Konya Technical University, Rectorate, Konya, Türkiye
0000-0002-5160-3210 ahmetakusta@hotmail.com

# References

Alam, M. S., Hossain, A. K. M. B., & Mohamed, F. (2022). Performance Evaluation of Recurrent Neural Networks for Indoor Camera Localization. *International Journal of Emerging Technology and Advanced Engineering*, *12*(8), 116–124. https://doi.org/10.46338/ijetae0822_15

Araci, D. T. (2019). FinBERT: Financial Sentiment Analysis with Pre-trained Language Models. Retrieved from https://arxiv.org/abs/1908.10063v1.

Bharathabau, K., Saurav, Vishal. (2023). Prediction and Sentiment Analysis of Stocks using Machine Learning. *International Journal For Science Technology And Engineering*, 11(5):6512-6519. doi: 10.22214/ijraset.2023.53169

Bouktif, S., Fiaz, A., and Awad, M. (2020). Augmented Textual Features Based Stock Market Prediction. *IEEE Access*, PP, 1. https://doi.org/10.1109/ACCESS.2020.2976725

Chai, T., & Draxler, R. R. (2014). Root mean square error (RMSE) or mean absolute error (MAE)? This is an argument against avoiding RMSE. *Geoscientific Model Development*, *7*(3), 1247–1250. https://doi.org/10.5194/gmd-7-1247-2014~

Chicco, D., Warrens, M. J., & Jurman, G. (2021). The coefficient of determination R-squared is more informative than SMAPE, MAE, MAPE, MSE, and RMSE in regression analysis evaluation. *PeerJ Computer Science*, *7*, 1–24. https://doi.org/10.7717/PEERJ-CS.623

Costola, M., Hinz, O., Nofer, M., & Pelizzon, L. (2023). Machine learning sentiment analysis of COVID-19 news and stock market reactions. *Research in International Business and Finance*, 64, 101881. https://doi.org/10.1016/J.RIBAF.2023.101881

Cui, H., ZhU, Y., Gu, F., & Wang, L. (2022). Research on stock price prediction using TextRank based text summarization technology and sentiment analysis. *2022 18th International Conference on Computational Intelligence and Security (CIS)*, 302-306. https://doi.org/10.1109/CIS58238.2022.00070

Gössi, S., Chen, Z., Kim, W., Bermeitinger, B., & Handschuh, S. (2023). FinBERT-FOMC: Fine-Tuned FinBERT Model with Sentiment Focus Method for Enhancing Sentiment Analysis of FOMC Minutes. *ICAIF 2023 - 4th ACM International Conference on AI in Finance*, 357–364. https://doi.org/10.1145/3604237.3626843

Gujjar, P. and Kumar, H. R. P. (2020). Opinion mining for the customer feedback using textbook. *Int J Sci Res Comput Sci Eng Inform Technol*, 72-76. https://doi.org/10.32628/CSEIT206418

Hajek, P., Novotny, J. and Kovarnik, J. (2022). Predicting Exchange Rate with FinBERT-Based Sentiment Analysis of Online News. *ACM International Conference Proceeding Series*, 133–138. https://doi.org/10.1145/3572647.3572667

Hodson, T. O. (2022). Root-mean-square error (RMSE) or mean absolute error (MAE): when to use them or not. *Geoscientific Model Development*, *15*(14), 5481–5487. https://doi.org/10.5194/gmd-15-5481-2022

Kaeley, H., Qiao, Y., & Bagherzadeh, N. (2023). Support for Stock Trend Prediction Using Transformers and Sentiment Analysis. *arXiv (Cornell University)*. https://doi.org/10.48550/arxiv.2305.14368

Kalyani, J.; Bharathi, Prof. H. N.; Jyothi, Prof. R. (2016). Stock trend prediction using news sentiment analysis. *International Journal of Computer Science and Information Technology*, 8(3), 67–76. https://doi.org/10.5121/ijcsit.2016.8306

Karimova L, Rakhmetulayeva S. (2023). Application of the Algorithm to Analyze Stock Prices Based on Sentiment Analysis. Cyst 2023-2023 *IEEE International Conference on Smart Information Systems and Technologies, Proceedings*, 214–220. https://doi.org/10.1109/SIST58284.2023.10223583

Kazemian, S., Zhao, S., & Penn, G. (2016). Evaluating Sentiment Analysis in the Context of Securities Trading. https://doi.org/10.18653/v1/P16-1197

Keen, P., Honnibal, M., Yankovsky, R., Karesh, D., Dempsey, E., Childs, W., Schnur, J., Qalieh, A., Ragnarsson, L., Coe, J. L., Calvo, A., Kulshrestha, N., Eslava, J., Albert, J., Harden,

Kovacs, Z., Kantor, D. B., & Fekete, A. (2008). Comparison of quantitative determination techniques with electronic tongue measurements. *American Society of Agricultural and Biological Engineers Annual International Meeting, 2008; ASABE 2008*, *11*, 6603–6615.

Kurniawan, F., Sulaiman, S., Konate, S., & Abdalla, M. A. A. (2023). Deep-Learning Approaches for MIMO Time-Series Analysis. *International Journal of Advances in Intelligent Informatics*, *9*(2), 286. https://doi.org/10.26555/ijain.v9i2.1092

Liu, J. X., Leu, J. S., & Holst, S. (2023). Stock price movement prediction based on Stocktwits investor sentiment using FinBERT and ensemble SVM. *PeerJ Computer Science*, 9, e1403. https://doi.org/10.7717/peerj-cs.1403

Liu, Q., Lee, W. S., Huang, M., & Wu, Q. (2023). Synergy between stock prices and investor sentiment in social media. *Borsa Istanbul Review*, 23(1), 76–92. https://doi.org/10.1016/j.bir.2022.09.006

Loria, S. (2013). *TextBlob: Simplified text processing* [GitHub repository]. GitHub. https://github.com/sloria/textblob

Ma, H., Ma, J., Wang, M. H., Li, P., & Du, W. (2021). A Comprehensive Review of Investor Sentiment Analysis in Stock Price Forecasting. https://doi.org/10.1109/ICISFall51598.2021.9627470

Pali, Kalpana, ve Laxmikant Tiwari. “Predictive Learning and Career Path Using Artificial Intelligence”. *2024 OPJU International Technology Conference (OTCON) on Smart Computing for Innovation and Advancement in Industry 4.0*, IEEE, 2024, ss. 1-9. *DOI.org (Crossref)*, https://doi.org/10.1109/OTCON60325.2024.10688178.

Palomino, M. A., Varma, A. P., Bedala, G. K., & Connelly, A. (2020). The Lack of Consensus Among Sentiment Analysis Tools. *Lecture Notes in Computer Science* (Including Subseries Lecture Notes in Artificial Intelligence and Lecture Notes in Bioinformatics), 12598 LNAI, 58–72. https://doi.org/10.1007/978-3-030-66527-2_5

Ridhawi, M. A., & Osman, H. A. (2023). We predict stock markets from sentiment and financial stock data using machine learning. *Proceedings of the Canadian Conference on Artificial Intelligence*. https://doi.org/10.21428/594757db.40c1a462

Robeson, S. M., and Willmott, C. J. (2023). Decomposition of the mean absolute error (MAE) into systematic and unsystematic components. *PLoS ONE*, *18*(2 February). https://doi.org/10.1371/journal.pone.0279774

Saxena, A., Jain, A., Sharma, P., Singla, S., & Ticku, A. (2023). Sentiment Analysis of Stocks Based on News Headlines Using NLP (pp. 124–135). https://doi.org/10.2991/978-94-6463-074-9_12

Scikit-learn: Machine Learning in Python, Pedregosa *et al.*, JMLR 12, pp. 2825-2830, 2011. From: https://scikit-learn.org/dev/modules/generated/sklearn.metrics.r2_score.html

Sharaf, M., Hemdan, E. E. D., El-Sayed, A., & El-Bahnasawy, N. A. (2023). An efficient hybrid stock trend prediction system during the COVID-19 pandemic based on stacked-LSTM and news sentiment analysis. *Multimedia Tools and Applications*, 82(16), 23945–23977. https://doi.org/10.1007/S11042-022-14216-W/TABLES/9

Štrimaitis, R., Stefanovič, P., Ramanauskaitė, S., & Slotkienė, A. (2021). Financial Context News Sentiment Analysis for the Lithuanian Language. *Applied Sciences, 11*(10), 4443. https://doi.org/10.3390/APP11104443

Varija, B., & Hegde, N. P. (2024). An Automated Analytics Framework for Stock Trend Analysis from Multi-Modal Data. *International Journal of Electronics and Communication Engineering*, Volume 11(1), 116–130. https://doi.org/10.14445/23488549/IJECE-V11I1P109

Wang, Y. (2023). Stock prediction using polyglot sentiment analysis on Twitter. *Journal of Student Research*, *12*(2). https://doi.org/10.47611/jsr.v12i2.1910

Willmott, C. J., & Matsuura, K. (2005). Advantages of mean absolute error (MAE) over the root mean square error (RMSE) in assessing average model performance. *Climate Research*, *30*(1), 79–82. https://doi.org/10.3354/cr030079

Willmott, C. J., Matsuura, K., & Robeson, S. M. (2009). Ambiguities inherent in sum-of-squares-based error statistics. *Atmospheric Environment*, *43*(3), 749–752. https://doi.org/10.1016/j.atmosenv.2008.10.005

Xiao, Q., & Ihnaini, B. (2023). Stock trend prediction using sentiment analysis. *PeerJ Computer Science*, *9*, e1293. https://doi.org/10.7717/peerj-cs.1293

Yang, M., Zhang, Y., & Ai, Y. (2023). Short-term Electricity Load Forecasting Based on Improved Sparrow Search Algorithm With Optimized BiLSTM. *Advanced Control for Applications Engineering and Industrial Systems*, *6*(2). https://doi.org/10.1002/adc2.160

Yang, Y., Christopher, M., UY, S., & Huang, A. (2020). FinBERT: A Pretrained Language Model for Financial Communications. In *Github*.

Yang, Yaqi, vd. “PRIME: Posterior Reconstruction of the Input for Model Explanations”. *Pattern Recognition Letters*, c. 176, Aralık 2023, ss. 202-08. *DOI.org (Crossref)*, https://doi.org/10.1016/j.patrec.2023.11.009.

Yekrangi, M., & Nikolov, N. S. (2023). Domain-Specific Sentiment Analysis: An Optimized Deep Learning Approach for the Financial Markets. *IEEE Access*, 11, 70248–70262. https://doi.org/10.1109/ACCESS.2023.3293733

Yeoh, E. D., Chung, T., & Wang, Y. (2023). Predicting price trends using sentiment analysis: A study of socialfi and gamezi cryptocurrencies. *Contemporary Mathematics*, 1089-1108. https://doi.org/10.37256/cm.4420232572

Zhang, H., Xu, J., Lei, L., Qiu, J., & Al-Shalabi, R. (2022). A Sentiment Analysis Method Based on Bidirectional Long Short-Term Memory Networks. *Applied Mathematics and Nonlinear Sciences*, *8*(1), 55–68. https://doi.org/10.2478/amns.2022.1.00015

Zuleaizal, S., Noor, H., Ibrahim, T. (2023). We associate deep learning and the news headlines sentiment for Bursa stock prices. *Indonesian Journal of Electrical Engineering and Computer Science*, 31(2):1041-1041. doi: https://doi.org/10.11591/ijeecs.v31.i2.pp1041-1049